\documentclass[prl,aps,twocolumn,floats,showpacs]{revtex4}

\usepackage{graphics}
\usepackage{amsmath}
\usepackage{citesort}
\usepackage{dcolumn}
\usepackage{epsfig}

\begin{document}

\title{
{\rm\small\hfill (submitted to Phys. Rev. Lett.)}\\
First-principles statistical mechanics study of the stability of a
sub-nanometer thin surface oxide in reactive environments: CO oxidation at Pd(100)}

\author{Jutta Rogal, Karsten Reuter, and Matthias Scheffler}

\affiliation{Fritz-Haber-Institut der Max-Planck-Gesellschaft,
Faradayweg 4-6, D-14195 Berlin, Germany}

\received{23rd August 2006}

\begin{abstract}
We employ a multiscale modeling approach to study the surface structure and composition of a Pd(100) model catalyst in reactive environments. Under gas phase conditions representative of technological CO oxidation ($\sim 1$\,atm, 300-600\,K) we find the system on the verge of either stabilizing sub-nanometer thin oxide structures or CO adlayers at the surface. Under steady-state operation this suggests the presence or continuous formation and reduction of oxidic patches at the surface, which could be key to understand the observable catalytic function.
\end{abstract}



\maketitle

Under realistic operating conditions heterogeneous catalysts are exposed to reactant pressures of the order of atmospheres. In recent years, there has been an increasing awareness that for oxidation catalysis, i.e. under atmospheric O$_2$ conditions, the surface of the employed transition metal (TM) catalysts might get oxidized. Instead of the pristine TM surface often studied in atomic detail under ultra-high vacuum (UHV) conditions, it could then be the surface oxide that only forms under reactive conditions which actuates the catalytic activity under atmospheric steady-state operation. Such situation induced by the reactive environment has recently been revealed for CO oxidation at so-called Ru catalysts, with two noteworthy conclusions \cite{over03,reuter03,reuter04b}: (i) There is a material's gap, i.e., in the initial period when the Ru metal is exposed to the reactive environment an oxide film is formed at the surface; and (ii) there is a pressure gap, i.e., on the once formed RuO$_2$ surface CO and O are only adsorbed in the proper ratio (necessary for efficient catalysis), if pressure and temperature of the reactants are sufficiently high.

As an element from the middle of the TM series Ru forms rather strong chemical bonds and RuO$_2$ is in fact the stable bulk phase under the reactive conditions \cite{reuter04a}. This is different for the late transition and noble metals, which are equally or even more frequently used in oxidation catalysis. In this study we therefore set out to further elucidate the situation by using CO oxidation over the Pd(100) model catalyst as an example. For this system, {\em in-situ} reactor scanning tunneling microscopy (STM) experiments had interpreted an observed roughening of the surface as oxide formation, accompanied with a notable increase in the catalytic activity \cite{hendriksen04}. The first-principles statistical mechanics calculations described in this Letter fully confirm this interpretation, in that indeed oxide formation in the reactive environment plays a key role also at the more noble Pd model catalyst. The striking gist is that this oxide is not a bulk-like film that once formed is stable and actuates the catalysis. Instead, our investigation points at the relevance of a sub-nanometer thin surface oxide film, which is likely continuously formed and reacted away in the on-going reaction.

Aiming at quantitative atomic-scale insight we employ a multiscale modeling approach to analyze the surface structure and composition of Pd(100), when exposed to a reactive gas phase consisting of O$_2$ and CO.  The atomic-level energetics of the system is obtained from density-functional theory (DFT) calculations, performed within the highly accurate full-potential augmented plane wave + local orbitals (LAPW/APW+lo) scheme \cite{wien2k}. The extended surface is described within a slab geometry and we carefully checked that the numerical uncertainty introduced by the finite basis set, as well as by the supercell parameters does not affect the reported results \cite{basis}. The only notable approximation with respect to the energetics underlying our approach is thus the approximate treatment of the exchange-correlation interaction, for which we use the generalized gradient approximation (GGA-PBE) \cite{perdew96}, and we will critically discuss this point below.

\begin{figure*} 
\epsfig{bbllx=0,bblly=0,bburx=575,bbury=272,clip=,
          file=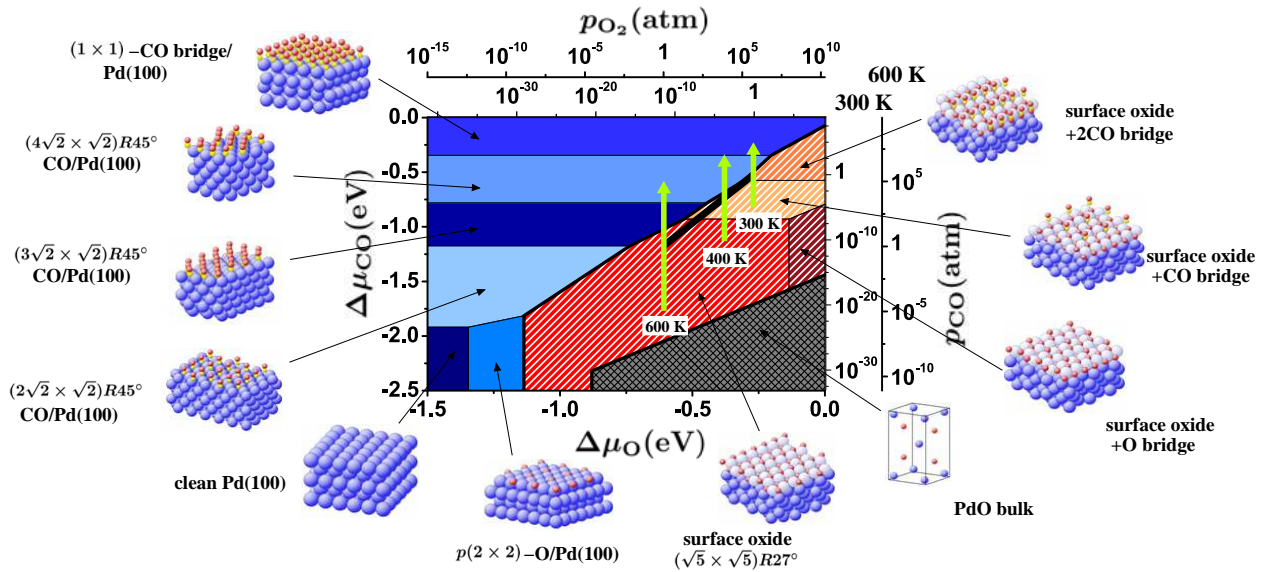,width=0.96\linewidth}
\caption{\label{fig1}
(Color online) Surface phase diagram for the Pd(100) surface in ``constrained thermodynamic equilibrium'' with an environment consisting of O$_2$ and CO. The atomic structures underlying the various stable (co-)adsorption phases on Pd(100) and the surface oxide, as well as a thick bulk-like oxide film (indicated by the bulk unit-cell), are also shown (Pd = large spheres, O = small spheres, C = white spheres). Phases involving the bulk oxide are cross-hatched, phases involving the surface oxide are hatched. The dependence on the chemical potentials of O$_2$ and CO in the gas phase is translated into pressure scales at 300\,K and 600\,K. The thick black line marks gas phase conditions representative of technological CO oxidation catalysis, i.e. partial pressures of 1\,atm and temperatures between 300-600\,K. The three vertical grey (green) lines correspond to the gas phase conditions employed in the kinetic Monte Carlo simulations shown in Fig. \ref{fig2} (see text).}
\end{figure*}

In a first step, we account for the effect of a finite gas phase environment within a constrained atomistic thermodynamics approach \cite{reuter03}. Here, the DFT energetics enters into the calculation of the Gibbs free energy of adsorption of possible states of the system. Neglecting kinetic effects of the on-going catalytic reactions on the surface structure and stoichiometry (at this stage), the surface is considered to be in equilibrium with $i$ separate reservoirs representing the $i$ gas phase species, each characterized by partial pressures $p_i$ and temperature $T$ (or more conveniently by a chemical potential $\Delta \mu_i(T,p_i)$ summarizing this two-dimensional dependence \cite{reuter03}). We verified that for the focus of this work, bulk and surface vibrational and configurational contributions to the Gibbs free energy of adsorption are not significant \cite{rogaltbp} and correspondingly neglect them here.

Within this approach we are then in a position to compare the stability of a wide range of possible surface structural models under ($T, p_{\textrm{O}_2}, p_{\textrm{CO}}$)-conditions ranging from UHV to atmospheric pressures and up to elevated temperatures. Aiming to assess a possible oxide formation at the surface, we consider bulk PdO to represent thick bulk-like oxide films \cite{reuter04a}. In addition we exploit the recent characterization of a $(\sqrt{5} \times \sqrt{5})R27^{\circ}$ (henceforth called $\sqrt{5}$) surface oxide structure, which corresponds to a sub-nanometer thin film of PdO(101) on the surface \cite{todorova03}. Including a large pool of combinatorially representative ordered (co-)adsorption phases of O and CO on this $\sqrt{5}$ surface oxide and on the pristine Pd(100) surface, we arrive at a set of 191 possible surface configurations. Out of these, we find only 11 to be most stable in a certain range of O$_2$ and CO chemical potentials, or equivalently ($T, p_{\textrm{O}_2}, p_{\textrm{CO}}$)-gas phase conditions. Figure \ref{fig1} summarizes these stability regions in $(\Delta \mu_{\rm O}, \Delta \mu_{\rm CO})$-space, where we additionally include the corresponding pressure scales for two representative temperatures.

A natural starting point to discuss the overall topology of this surface phase diagram is the bottom left part of Fig. \ref{fig1}. This corresponds to vanishing concentrations of gas phase species and correspondingly, the clean Pd(100) surface results as the most stable system state. Moving from this situation to the right in the surface phase diagram, e.g. by increasing the O$_2$ pressure while keeping the CO pressure low, we find surface structures containing an increasing amount of oxygen. These are a $p(2 \times 2)$ adsorption phase with O atoms adsorbed at on-surface hollow sites at a coverage $\Theta = 0.25$\,monolayer (ML), the $\sqrt{5}$ surface oxide ($\Theta = 0.8$\,ML) and bulk-like thick oxide films; a sequence that was recently confirmed in {\em in-situ} surface X-ray diffraction measurements \cite{lundgren04}. In the complementary case of low oxygen pressure and increasing CO pressure (moving from the bottom left in Fig. \ref{fig1} upwards) a series of ordered CO adsorption phases on Pd(100) are stabilized, most of which have already been characterized in UHV experiments \cite{bradshaw78}. For simultaneously increased O$_2$ and CO gas concentrations, i.e. moving now up right along the diagonal in Fig. \ref{fig1}, there are three phases with O or CO adsorbed at the $\sqrt{5}$ surface oxide structure, but interestingly no co-adsorption structures on Pd(100). Although our set of surface configurations compared in the thermodynamic approach comprises also a large number of such structures, we find them largely disfavored by overall repulsive lateral interactions among the adsorbates, which is consistent with the experimentally observed tendency of adsorbed O and CO to form separate domains on the Pd(100) surface \cite{stuve84}.

The overall topology of the constrained surface phase diagram can thus be divided into three different parts. First, a region of gas phase conditions, where bulk-like thick oxide films are stable (cross-hatched area). Second, a region comprising various adsorption phases on the $\sqrt{5}$ surface oxide (hatched area), and finally, a region with different adsorption phases on Pd(100). Particularly intriguing is that gas phase conditions representative of technological CO oxidation catalysis ($p_i \sim$\,1\,atm, $T \sim 300-600$\,K) fall almost exactly on the boundary between the stability region of adsorption phases on the $\sqrt{5}$ surface oxide and the stability region of CO-covered Pd(100) phases, cf. Fig. \ref{fig1}. On the basis of these results we would correspondingly be able to draw two important conclusions with respect to the relevance of oxide formation in the reactive environment: (i) The formation of a thick bulk-like oxide film at the surface as in the case of the Ru catalyst can be ruled out for Pd(100) under any gas phase conditions representative of technological CO oxidation catalysis. (ii) The stability region of the recently characterized $\sqrt{5}$ surface oxide, on the other hand, does extend to such conditions.

Before further discussing the implications of these findings, it is appropriate to scrutinize the major approximation of our approach, namely the xc functional employed in the DFT calculations. To this end, we recomputed Fig. \ref{fig1} using the local-density (LDA \cite{perdew92}) and another gradient corrected (GGA-rPBE \cite{hammer99}) xc functional, which both yield notably different O$_2$ and CO gas phase binding energies compared to the GGA-PBE \cite{hammer99}. In both cases we obtain the same topology of the surface phase diagram, i.e. we find the same stable phases in the same regions as in Fig. \ref{fig1}. However, in general there are notable shifts in the location of the boundaries between the various phases in $(\Delta \mu_{\rm O}, \Delta \mu_{\rm CO})$-space. Strikingly, this does not apply to the boundary between the CO-covered Pd(100) phases and the adsorption phases on the $\sqrt{5}$ surface oxide. For all three xc functionals, the catalytically relevant gas phase conditions lie very close to this boundary, but still within the stability region of the surface oxide. This situation is neither changed, when approximately accounting for the bulk and surface vibrational contributions to the Gibbs free energy of adsorption \cite{reuter03,rogaltbp}. Considering the configurational entropy contributions would smear out the infinitely sharp boundary drawn in Fig. \ref{fig1} over a small, but finite range of chemical potentials. While the width of this phase coexistence range can be estimated \cite{reuter03}, we defer its discussion to the kinetic simulations below, where such configurational effects are fully included.

The constrained thermodynamics employed up to here is understood as a starting point, helping us to identify the regions in $(T,p_i)$-space that deserve closer attention by a higher-level approach considering chemical reactions and kinetics. In a second step, we therefore explicitly account for such effects by performing first-principles kinetic Monte Carlo (kMC) simulations to specifically investigate the stability range of the $\sqrt{5}$ surface oxide. In these simulations we follow the time evolution of the system as described by a master equation and use rates for the individual elementary processes that are obtained from the DFT GGA-PBE energetics together with transition-state theory \cite{reuter04b}. We focus here on the lattice structure of the $\sqrt{5}$ surface oxide, in particular on the surface kinetics involving the hollow and bridge sites. We consider all non-concerted adsorption, desorption, diffusion and Langmuir-Hinshelwood reaction processes involving these
 sites. Dissociative O$_2$ adsorption requires two neighboring sites, and is only hindered by a computed dissociation barrier of 1.9\,eV in the case of adsorption into two bridge sites. CO adsorption is unimolecular and not hindered by adsorption barriers. Nearest-neighbor lateral interactions are taken into account in the elementary process rates and we verified that the kMC set-up reproduces the part of the thermodynamic phase diagram involving all $\sqrt{5}$ phases (hatched part of Fig. \ref{fig1}), when the reaction events are not allowed to occur \cite{rogaltbp}.

The kMC simulations were performed for fixed $(T, p_{\rm O_2}, p_{\rm CO})$-conditions on a lattice comprising $(50 \times 50)$ surface unit-cells, evaluating the average occupation at the various surface sites after steady-state was reached. We carried out simulations in the temperature range 300-600\,K and for fixed $p_{\rm O_2} = 1$\,atm. With an initially set low CO partial pressure, we are then in the middle of the thermodynamic stability region of the $\sqrt{5}$ surface oxide shown in Fig. \ref{fig1}. And indeed, we exactly obtained average occupations pertinent to the geometric structure of the corresponding phases in Fig. \ref{fig1}, namely two on-surface oxygen atoms sitting in their preferred hollow sites \cite{todorova03}, while the bridge sites are either empty or filled with one or two CO molecules or one O atom per surface unit-cell. Under such gas phase conditions, kinetic effects due to the on-going reactions are thus negligible and we simply reproduce the results obtained within the atomistic thermodynamic approach.

\begin{figure} 
\epsfig{bbllx=26,bblly=30,bburx=589,bbury=306,clip=,
          file=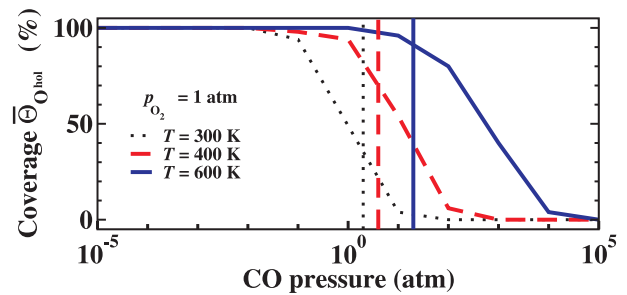,width=0.96\columnwidth}
\caption{\label{fig2}
Kinetic Monte Carlo (kMC) results for the stability of the $\sqrt{5}$ surface oxide. Shown is the average occupation $\overline{\Theta}_{\rm O^{\rm hol}}$ of the hollow sites with oxygen in the $\sqrt{5}$ lattice structure. The chosen gas phase conditions ($T=300, 400, 600$\,K, $p_{\rm O_2} = 1$\,atm, variable CO pressure) correspond to the three vertical lines drawn in Fig. \ref{fig1}. In the intact $\sqrt{5}$ surface oxide structure $\overline{\Theta}_{\rm O^{\rm hol}} = 100$\,\%, and the reduction obtained in the kMC simulations occurs at CO partial pressures quite close to those corresponding to the stability boundary as obtained in the constrained thermodynamic approach (transition from the hatched to the plain areas in Fig. \ref{fig1}), marked here by the corresponding vertical lines.}
\end{figure}

In subsequent simulations we then gradually increased the CO partial pressure, which brings us closer and closer to the boundary of the stability region of the $\sqrt{5}$ surface oxide as indicated by the vertical lines in Fig. \ref{fig1}. The most important surface reactions that can occur between the adsorbed species under these gas phase conditions are ${\rm O}^{\rm hollow}+ {\rm CO}^{\rm bridge} \rightarrow {\rm CO_2}$ with a calculated reaction barrier of $\Delta E^{\mathrm{reac}}_{\mathrm{O^{hol} + CO^{br}}} = 0.9$\,eV and ${\rm O}^{\rm bridge} + {\rm CO}^{\rm hollow} \rightarrow {\rm CO_2}$ with $\Delta E^{\mathrm{reac}}_{\mathrm{O^{\rm br} + CO^{\rm hol}}} = 0.5$\,eV. Kinetic effects when these reactions consume surface species faster than they can be replenished from the gas phase may lead to the destabilization of the $\sqrt{5}$ surface oxide at already lower $p_{\rm CO}$ than predicted by the stability boundary in the thermodynamic phase diagram. To determine the onset of this structural decomposition in the kMC simulations we focus on the average occupation $\overline{\Theta}_{\rm O^{\rm hol}}$ of O species at the hollow sites. As just discussed, these sites are always fully occupied in the intact surface oxide structure and we interpret a $\overline{\Theta}_{\rm O^{\rm hol}}$ reduced to 80\% as a sign for the onset of the structural decomposition. The corresponding data as a function of the CO partial pressure for various temperatures is shown in Fig. \ref{fig2}.

The depopulation of the hollow sites occurs at all temperatures over a certain range of CO partial pressures. The deduced critical pressures for the onset of oxide decomposition depend nevertheless little on the present choice of reduction to 80\,\% occupation, and we also verified they would be little affected by slight 0.1\,eV uncertainties in the employed DFT GGA-PBE surface reaction barriers. As apparent from Fig. \ref{fig2} the stability range of the $\sqrt{5}$ surface oxide defined by these critical pressures is in all cases quite similar to the one obtained within the constrained atomistic thermodynamic approach. Still, the small variation of the stability region induced by the reaction kinetics has a notable effect. Whereas in the constrained equilibrium approach the catalytically relevant environments ($p_i = 1$\,atm, 300-600\,K, black line in Fig. \ref{fig1}) were always fully within the stability region of the $\sqrt{5}$ surface oxide, this is no longer the case when the reaction kinetics are taken into account. At the lower temperatures considered, slightly oxygen-rich conditions are now required to stabilize the surface oxide film. At 400\,K, this is a partial pressure ratio of $p_{\rm O_2}/p_{\rm CO} \approx 10:1$, which is in good agreement with the recent {\em in-situ} reactor STM experiments \cite{hendriksen04} performed at such gas phase conditions. At the higher temperatures, the stability of the surface oxide is enhanced, and our simulation results predict that at 600\,K, the surface oxide is still stable even up to a pressure ratio of 1:1. This suggests that at least at these higher temperatures the catalytically most relevant gas phase conditions correspond to a transient situation where either sub-nanometer thin oxide structures or CO adlayers are just stabilized at the surface, such that small changes (or even fluctuations) in the partial pressures may have profound consequences on the global or local surface composition and structure.

Employing DFT together with constrained thermodynamics as well as statistical mechanics we therefore provide evidence for the relevance of oxide formation at the Pd(100) model catalyst under gas phase conditions representative of technological CO oxidation. Depending on temperature, the system is on the verge of stabilizing sub-nanometer thin oxide structures or CO adlayers at the surface. Similar to the case of CO oxidation at Ru catalysts, it is therefore again a fluctuation-rich region close to boundaries in the computed surface phase diagram that is of particular interest for the catalytic function. While in Ru the coexistence is between different adsorption phases on an otherwise stable oxide film, at the more noble Pd(100) it involves the presence of CO on pristine Pd and CO on surface oxide patches. Under steady-state operation this suggests possible oscillations between these two states, which could then also drive a spatio-temporal pattern formation.

The EU is acknowledged for financial support under contract NMP3-CT-2003-505670 (NANO$_2$), and the DFG for support within the priority program SPP1091.


\begin{thebibliography}{99}
\bibitem{basis}
LAPW/APW+lo basis set parameters: $R_{\rm{MT}}^{\rm{Pd}}=$2.0\,bohr,
$R_{\rm{MT}}^{\rm{O}}=$1.0\,bohr, $R_{\rm{MT}}^{\rm{C}}=$1.0\,bohr, $E^{\rm max}_{\rm wf} = 20$\,Ry, $E^{\rm max}_{\rm pot} = 196$\,Ry. A $(10 \times 10 \times 1)$ Monkhorst-Pack grid was used for $(1 \times 1)$-Pd(100), and reduced in larger surface unit-cells to maintain the same reciprocal space sampling. The supercells were composed of at least 14\,{\AA} vacuum and 5 Pd(100) layer slabs, with adsorption of O/CO or the surface oxide plus O/CO at both sides. With this computational setup, the relative stabilities of different surface phases are converged to within $\pm 5$\,meV/{\AA}$^2$.
\end{thebibliography}
\end{document}